\documentclass[preprint,12pt]{elsarticle}

\usepackage{times}
\usepackage{soul}
\usepackage{xurl}
\usepackage[switch]{lineno}

\usepackage{graphicx} 
\usepackage{subcaption}
\usepackage{nomencl}
\usepackage{etoolbox}
\usepackage{amssymb}
\usepackage{hyperref}
\usepackage[utf8]{inputenc}
\usepackage[T1]{fontenc}
\usepackage{float}
\usepackage{color}
\usepackage{lineno}
\usepackage{amsmath}
\usepackage[linesnumbered]{algorithm2e}
\usepackage{caption}
\usepackage{booktabs}
\usepackage{array}
\usepackage{pgfplots}
\usepackage[euler-digits]{eulervm}
\usepackage[eulergreek]{sansmath}
\usepackage{amsmath} 
\usepackage{amsthm}

\usepackage{algorithmic}
\usepackage{amsfonts}
\usepackage{subcaption}

\usepackage{float}
\usepackage[hypcap]{caption}
\usepackage{subcaption}
\usepackage{algorithm2e}

\usepackage{mathtools}
\DeclarePairedDelimiter\floor{\lfloor}{\rfloor}
\usepackage[shortlabels]{enumitem}

\usepackage{listings}

\lstdefinestyle{custommatlab}{
  belowcaptionskip=1\baselineskip,
  breaklines=true,
  frame=L,
  xleftmargin=\parindent,
  language=python,
  showstringspaces=false,
  basicstyle=\footnotesize\ttfamily,
  keywordstyle=\bfseries\color{green!40!black},
  commentstyle=\itshape\color{purple!40!black},
  identifierstyle=\color{blue},
  stringstyle=\color{orange},
}
\usetikzlibrary{positioning}
\pgfplotsset{compat=1.18}
\newlength\algowd

\newcommand{\revision}[1]{\textcolor{black}{#1}}

\begin{document}

\begin{frontmatter}

\title{Matrix-by-matrix multiplication algorithm with   $O(N^2log_2N)$  computational cost for variable precision arithmetic}

\author{Maciej Paszy\'nski}

\address{AGH University of Krakow,
Poland \\
e-mail:  maciej.paszynski@agh.edu.pl \\
home.agh.edu.pl/paszynsk}

\begin{abstract}
In this paper, we show that assuming the availability of the processor with variable precision arithmetic, we can compute matrix-by-matrix multiplications in  $O(N^2log_2N)$ computational complexity. Namely, we replace the standard matrix-by-matrix multiplications algorithm proposed by Jacques Philippe Marie Bin\'et reading
\begin{equation}
\begin{bmatrix} A_{11} & A_{12} \\ A_{21} & A_{22} \end{bmatrix}
\begin{bmatrix} B_{11} & B_{12} \\ B_{21} & B_{22} \end{bmatrix} = \begin{bmatrix} A_{11}B_{11}+A_{12}B_{21} & A_{11}B_{12}+A_{12}B_{22} \\
A_{21}B_{11}+A_{22}B_{21} & A_{21}B_{12}+A_{22}B_{22}  \end{bmatrix}  \notag
\end{equation}
by the following
\begin{eqnarray}
\begin{bmatrix} A_{11} & A_{12} \\ A_{21} & A_{22} \end{bmatrix}
\begin{bmatrix} B_{11} & B_{12} \\ B_{21} & B_{22} \end{bmatrix} = \notag \\
\Bigl\lfloor\begin{bmatrix} (A_{11}+\epsilon A_{12})(B_{11}+1/{\epsilon}B_{21})  &
(A_{11}+\epsilon A_{12})(B_{12}+1/{\epsilon}B_{22}) \\
(A_{21}+\epsilon A_{22})(B_{11}+1/{\epsilon}B_{21}) &
(A_{21}+\epsilon A_{22})(B_{12}+1/{\epsilon}B_{22})   \end{bmatrix} \Bigr\rfloor \% \frac{1}{\epsilon} \notag
\end{eqnarray}
where $\lfloor \rfloor$ denotes the floor, and $\%$ denotes the modulo operators.
We reduce the number of block matrix-by-matrix multiplications from 8 to 4,
keeping the number of additions equal to 4, and additionally introducing 4 multiplications of a block matrices by $\epsilon$ or $\frac{1}{\epsilon}$, and 4 floor and 4 modulo operations.
The resulting computational complexity for two matrices of size $N\times N$ can be estimated from recursive equation $T(N)=4(N/2)^2$ (multiplication of a matrix by $\epsilon$ and $1/\epsilon$) plus $4(N/2)^2$ (additions of two matrices) plus $2N^2$ (floor and modulo) plus $4T(N/2)$ (four recursive calls) as $O(N^2log_2N)$.
These multiplications of blocks of a matrix by number scales like $O((N/2)^2)$.
In other words, while having a processor that can compute multiplications, additions, modulo and floor operations with variable precision arithmetic in $O(1)$, we can obtain a matrix-by-matrix multiplication algorithm with $O(N^2log_2N)$ computational complexity, which is bounded by $O(N^{2+\delta})$, for arbitrary small $\delta>0$, e.g. $O(N^{2.001})$ .
We also present a MATLAB code using \emph{vpa} variable precision arithmetic emulator that can multiply matrices of size $N\times N$ using $(4log_2N+1)N^2$ variable precision arithmetic operations. This emulator uses $O(N)$ digits to run our algorithm.

\end{abstract}
	

\end{frontmatter}

\section{Introduction}
The matrix multiplication algorithm is a core element of scientific computing and machine learning. Its computational complexity affects other algebraic algorithms \cite{c22}, including the matrix inversion (expressed in a recursive way), LU factorization (extension of Gaussian elimination), the determinant computations (via multiplications of diagonals of L and U), as well as neural-network training.
The computational complexity of traditional matrix-by-matrix multiplications introduced by Jacques Philippe Marie Bin\'et in 1812 \cite{c14} is $O(N^3)$. It has been reduced by Volker Strassen to $O(N^{log_2(7)}=N^{2.8074})$ \cite{c20}
 and further by Coppersmith-Winegrod \cite{c8} down to $O(N^{2.496})$.
Since then, there has been a long series of papers improving the upper bound down to $O(N^{2.37286})$
\cite{c1,c3,c4,c5,c6,c7,c9,c10,c13,c15,c16,c17,c18,c21,c23}.
Recently, AlphTensor AI tools with reinforcement learning discovered several generalizations of the Strassen algorithm for different dimensions of matrix blocks \cite{Nature}. 

In this paper, \revision{we show that assuming the availability of the processor with long precision arithmetics,} we propose an algorithm that delivers $O(N^2log_2N)$ computational complexity of matrix-by-matrix multiplication. While designing our algorithm, we made the following assumptions:
\begin{enumerate}[(a)]
\item \revision{The modern processors allow for ${\cal O}(1)$ processing of numbers up to 512 bits (see for example AVX-512, the 512-bit extensions to the 256-bit Advanced Vector Extensions SIMD instructions for x86 instruction set architecture (ISA) proposed by Intel in July 2013, and first implemented in the 2016 Intel Xeon Phi x200 and then later in a number of AMD and other Intel CPUs). Using this architecture, it is possible to process small matrices in ${\cal O}(1)$ time. Further extension of long precision arithmetic has to implement fast algorithms for multiplication of long integers. The fastest known is the Harvey–van der Hoeven ${\cal O}(d\log d)$ algorithm \cite{Hoeven}, faster than the classical Karatsuba ${\cal O}(d^{1.58})$ algorithm \cite{c12}. This is the case if the processors can process the long integers using a single core.}
\revision{Having multicore processing units, the long arithmetics numbers can be indeed processed in  ${\cal O}(\log d )$, using, e.g., FFT-based multiplication algorithms (like Schönhage-Strassen \cite{SS} or Fürer's algorithm \cite{Furer}).}
\revision{Our algorithm uses integers up to size $N$, where $N$ is the matrix size. By using the multicore processing unit, we can implement our algorithm in ${\cal O}(N^2 \log N)$ computational complexity.}
\item We will also assume that the computational complexity of modulo and floor operation executed on a number with a variable number of digits is $O(1)$.
\item We are estimating the computational complexity in terms of the big O notation 
as well as the exact computational cost.
\item We will assume that we have immediate access to data and that the cost of accessing matrix entries is zero. 
\end{enumerate}

\revision{Discovering the matrix multiplication algorithm with ${\cal O}(N^2 \log N)$ computational complexity will have practical implications only if the computational cost $CN^2\log N$ has a small constant $C$.
The numerical algebra libraries \cite{c2} employ the Bin\'et algorithm with ${\cal O}(N^3)$ computational complexity because its exact computational cost is low (it is like $2N^3-N^2$). The Strassen algorithm is like $N^{\log_2(7)}$ multiplications plus $6(N^2-1)$ additions/subtraction. Hence, it is still possible to efficiently use for small matrices, and the examples of that are described in \cite{c19,c11} (using sequential and parallel implementations).
However, to our best knowledge, the algorithms discovered after the Strassen method have never been implemented in a linear algebra library.
For example, the Coppersmith-Winograd algorithm \cite{c7} is considered a galactic algorithm due to the huge constant in front of the computational cost $C\cdot N^{2.3728596}$.
We are not aware of the exact estimate of the constant, but Wirginia Vassilevska-Williams in \cite{Williams1} acknowledges that these algorithms are of no practical use due to extremely large constants and numerical instability.
Additionally, Alman and Vassilewska-Williams in \cite{Williams2} acknowledge that despite these asymptotic improvements, all current algorithms are galactic, and none are faster in practice than Strassen's algorithm.}

\revision{The exact computational cost of the proposed algorithm is $4N^2\log N+N^2$. The constant is equal to 4.
The algorithm requires long-precision arithmetics, which is possible to develop using actual technology.
It will work in $N^2(4\log N+1)$ time for small matrices, providing we have access to long precision arithmetics.
The best algorithms for fast multiplications of the long integers have ${\cal O}(d \log d)$ \cite{Hoeven} computational complexity, where d is the number of digits.
Having multicore processing units, the long arithmetics numbers can be indeed processed in ${\cal O}(\log d)$,
using a concurrent version of e.g. FFT-based multiplication algorithms (like Schönhage–Strassen \cite{SS} or Fürer's algorithm \cite{Furer}) or Harvey–van der Hoeven ${\cal O}(d\log d)$ \cite{Hoeven}.}

\revision{The computational complexity of the matrix-matrix multiplication algorithm for integers is identical to the computational complexity for matrix-matrix multiplication with fixed-point arithmetics (non-integers with a fixed number of digits before and after the dot). Fixed-point number operations asymptotically have the same cost as single-digit integers assuming fixed-size arithmetic since matrix multiplication can be decomposed as
\begin{equation}
AB=(10^0A_0+10^1A_1+...+10^dA_d)(10^0B_0+10^1B_1+...+10^dB_d)
\end{equation}
For example, just for two digits, we have $d=2$
\begin{equation}
(10^0A_0+10^1A_1)(10^0B_0+10^1B_1)=
10^0A_0B_0+10^1A_0B_1+10^1A_1B_0+10^2A_1B_1
\end{equation}
so we have
$d^2=2^2=4$ terms.
The multiplication of matrices can also be decomposed into
\begin{equation}
\begin{aligned}
AB=& (10^{-d}A_{-d}+...+10^{-1}A_{-1} +10^0A_0+10^1A_1+...+10^dA_d)  \\ &(10^{-d}B_{-d}+...+10^{-1}B_{-1} +10^0B_0+10^1B_1+...+10^dB_d)
\end{aligned}
\end{equation}
Using this direct decomposition, we can express the multiplication of matrices with long integers or fixed-point numbers (the fixed number of digits before and after the decimal point) as $d^2$ multiplication of matrices with single digits.
It is also possible to apply the Karatsuba algorithm \cite{c12} for multiplication of decomposed matrices,
which implies $d^{1.58}$ multiplications of matrices with single digits for long integers, and $4d^{1.58}$ multiplication of matrices with single digits for fixed point arithmetics.
Thus, the computational complexity of the algorithm processing non-integers is identical, and the computational cost is multiplied by a factor of $4d^{1.58}$.
It is also possible to apply Harvey, der Hoeven algorithm \cite{Hoeven} for multiplication of decomposed matrices, which implies O(dlogd) multiplications of matrices with single digits or non-integers.}

\section{Key idea of our algorithm}

Let us assume that we intend to multiply two vectors $x = \begin{bmatrix} a & b \end{bmatrix} \cdot \begin{bmatrix}c & d \end{bmatrix} = ac + bd$.
Instead of this classical method, we employ the following equality
\begin{equation}
\begin{aligned}
x& =\lfloor (a+ \epsilon b) (c+1/ \epsilon d) \rfloor  \% \frac{1}{\epsilon}  \\
& = \lfloor (ac+ 1/\epsilon ad + \epsilon bc + bd) \rfloor  \% \frac{1}{\epsilon}  \\
& = (ac+bd)
\end{aligned}
\end{equation}
The term $\epsilon bc$ is of order $\epsilon$ and it is considered to be too small. The term $1/\epsilon ad$ is of order $1/\epsilon$ and it is considered to be too big. We cancel them out \emph{after} performing the multiplication $ (a+ \epsilon b) (c+1/ \epsilon d) $. We cancel small term using \emph{floor} operator and we cancel large term using \emph{modulo} operator. We cancel them \emph{after} performing the multiplication, so the term $ \epsilon b$ can multiply the term $1/\epsilon d$, cancel out $\epsilon$ with $1/\epsilon$ and remain $bd$. 
In the general case, we replace $a, b, c, d$ with matrices and use recursion.
Note that in this key formula, we have three multiplications ($\epsilon b$, $\frac{1}{\epsilon} d$ and $(a+ \epsilon b) (c+1/ \epsilon d)$). Still, two of them are the multiplications by numbers ($\epsilon$ and $\frac{1}{\epsilon}$).
The cost of multiplication of a matrix of size $N\times N$ by a number is $O(N^2)$. The cost of addition of two matrices of size $N\times N$ is also $O(N^2)$.
The cost of multiplication of two matrices of size $N\times N$ is $O(N^3)$, so we have reduced the number matrix-by-matrix multiplications from two to one. Additionally, the cost of modulo or floor operation of matrix entries is $O(N^2)$ for a matrix of size $N \times N$.

\section{The generalization into matrices}

Instead of recursive Bin\'et method

\begin{equation}
\begin{bmatrix}
A_{11} & A_{12} \\ 
A_{21} & A_{22} 
\end{bmatrix}
\begin{bmatrix}
B_{11} & B_{12} \\
B_{21} & B_{22}
\end{bmatrix}=
\end{equation}
\begin{equation}
\begin{bmatrix}
A_{11}B_{11}+A_{12}B_{21}  & A_{11}B_{12}+A_{12}B_{22} \\
A_{21}B_{11}+A_{22}B_{21}  & A_{21}B_{12}+A_{22}B_{22} \\
\end{bmatrix}
\end{equation}

We compute

\begin{equation}
\begin{bmatrix}
A_{11} & A_{12} \\ 
A_{21} & A_{22} 
\end{bmatrix}
\begin{bmatrix}
B_{11} & B_{12} \\
B_{21} & B_{22}
\end{bmatrix}=
\end{equation}
\begin{equation}
\Bigl\lfloor\begin{bmatrix} (A_{11}+\epsilon A_{12})(B_{11}+1/{\epsilon}B_{21})  & 
(A_{11}+\epsilon A_{12})(B_{12}+1/{\epsilon}B_{22}) \\ 
(A_{21}+\epsilon A_{22})(B_{11}+1/{\epsilon}B_{21}) &
(A_{21}+\epsilon A_{22})(B_{12}+1/{\epsilon}B_{22})   \end{bmatrix} \Bigr\rfloor \% \frac{1}{\epsilon} 
\end{equation}

The multiplication of matrices are performed in the recursive way. 
The pseudocode is summarized below.

\begin{enumerate}
\item [M] = MM(A,B,k)
\item Select proper $\epsilon$ and precision 
\item
$A = \begin{bmatrix} A_{11} & A_{12} \\ A_{21} & A_{22} \end{bmatrix}$
\item 
$B = \begin{bmatrix} B_{11} & B_{12} \\ B_{21} & B_{22} \end{bmatrix}$
\item 
$R_1 = A_{11}+\epsilon A_{12}$
\item 
$R_2 = A_{21}+\epsilon A_{22}$
\item
$C_1 = B_{11}+1/\epsilon B_{21}$
\item
$C_2 = B_{12}+1/\epsilon B_{22}$
\item
if $N/2 > 1$

  $\; \; M_{11} = MM(R_1,C_1,k+1);$

  $\; \; M_{12} = MM(R_1,C_2,k+1);$

  $\; \; M_{21} = MM(R_2,C_1,k+1);$

  $\; \; M_{22} = MM(R_2,C_2,k+1);$
\item
else

  $\; \; M_{11} = R_1 C_1$

  $\; \; M_{12} = R_1 C_2$

  $\; \; M_{21} = R_2 C_1$

  $\; \; M_{22} = R_2 C_2$

\item
$M = \begin{bmatrix} M_{11} & M_{12} \\ M_{21} & M_{22} \end{bmatrix}$

\item
$M = \lfloor M \rfloor \% precision$

\item return $M$
\end{enumerate}

We have replaced 8 multiplications of $(N/2)\times(N/2)$ blocks with 4 multiplications (and classical computational complexity of matrix-by-matrix multiplication of matrices of size $(N/2)\times(N/2)$ is $O((N/2)^3)$.
We still have 4 additions of $(N/2)\times(N/2)$ blocks (and classical computational complexity of addition of two matrices of size $(N/2)\times(N/2)$ is $O((N/2)^2)$.
We have also introduced 2 multiplications of $(N/2)\times(N/2)$ matrices by $\epsilon$,
2 multiplications of $(N/2)\times(N/2)$ matrices by $1/\epsilon$, and the modulo and floor operation of entries of $N\times N$ matrix.
\section{Estimation of the computational complexity}

To estimate the computational complexity we need to solve the following recursive equation
\begin{equation}
\begin{aligned}
T(N)&=\underbrace{4(N/2)^2}_{\textrm{multiplication of a matrix by $\epsilon$ and $1/\epsilon$}} \\&+ \underbrace{4(N/2)^2}_{\textrm{additions of two matrices}}+ \underbrace{4T(N/2)}_{\textrm{recursive calls}} + \underbrace{2N^2}_{\textrm{floor and modulo}}
\end{aligned}
\end{equation}
The solution to this recursive equation delivers $O(N^2log_2N)$ computational complexity, which is bounded by $O(N^{2+\delta})$, for arbitrary small $\delta>0$, e.g. $O(N^{2.001})$  which is lower than the lowest known computational complexity $O(N^{2.37286})$ of matrix-by-matrix multiplications.

\section{Exemplary MATLAB implementation}

We present the exemplary MATLAB code using \emph{vpa} \cite{vpa} variable precision arithmetic emulator that can multiply matrices with $O(N^2log_2N)$ computational cost when counting operations performed by variable precision arithmetic engine.
We hope that in the future, the emulator of variable precision arithmetic could be replaced by a real processor.

\begin{lstlisting}
function [C,count] = MMvpa(A,B,k,count)
sz = size(A); N=sz(1);
base=4;
epsilon_base=vpa(power(10,-base));
epsilon = vpa(epsilon_base); prec=2*base;
if k>1
  for i=2:k
    epsilon=epsilon*epsilon; prec=prec*2;
  end
end
Nhalf = N/2;
A11 = vpa(A(1:Nhalf,1:Nhalf));
A12 = vpa(A(1:Nhalf,Nhalf+1:N));
A21 = vpa(A(Nhalf+1:N,1:Nhalf));
A22 = vpa(A(Nhalf+1:N,Nhalf+1:N));
B11 = vpa(B(1:Nhalf,1:Nhalf));
B12 = vpa(B(1:Nhalf,Nhalf+1:N));
B21 = vpa(B(Nhalf+1:N,1:Nhalf));
B22 = vpa(B(Nhalf+1:N,Nhalf+1:N));
row1 = (A11+epsilon*A12); count = count + 2*Nhalf*Nhalf;
row2 = (A21+epsilon*A22); count = count + 2*Nhalf*Nhalf;
col1 = (B11+1.0/epsilon*B21); count = count + 2*Nhalf*Nhalf;
col2 = (B12+1.0/epsilon*B22); count = count + 2*Nhalf*Nhalf;
if Nhalf > 1
  [C11,count] = MMvpa(row1,col1,k+1,count);
  [C12,count] = MMvpa(row1,col2,k+1,count);
  [C21,count] = MMvpa(row2,col1,k+1,count);
  [C22,count] = MMvpa(row2,col2,k+1,count);
else
  C11 = row1*col1; count+1;
  C12 = row1*col2; count+1;
  C21 = row2*col1; count+1;
  C22 = row2*col2; count+1;
end
C=vpa(zeros(N,N),prec);
C(1:Nhalf,1:Nhalf)=C11(1:Nhalf,1:Nhalf);
C(1:Nhalf,Nhalf+1:N)=C12(1:Nhalf,1:Nhalf);
C(Nhalf+1:N,1:Nhalf)=C21(1:Nhalf,1:Nhalf);
C(Nhalf+1:N,Nhalf+1:N)=C22(1:Nhalf,1:Nhalf);
C = mod(floor(C),1/epsilon_base); count+2*N*N;
return
\end{lstlisting}

Using default 32 digits of variable precision arithmetic we can multiply matrices of dimensions $8\times 8$.
For example, generating two matrices with digits

\begin{lstlisting}
A=floor(rand(8,8)*10)
A =
     8     8     5     0     0     4     8     0
     4     8     6     5     0     3     7     5
     2     1     6     7     3     3     7     1
     2     7     3     7     0     3     1     4
     2     1     8     2     7     7     6     0
     5     4     5     7     2     5     9     8
     6     6     4     5     7     6     4     7
     3     0     0     0     0     2     8     9
B=floor(rand(8,8)*10)
B =
     4     1     4     4     6     3     2     2
     2     5     8     1     7     6     7     6
     4     2     8     9     9     7     1     4
     6     3     6     9     0     5     1     7
     1     5     6     5     2     1     3     5
     0     0     5     4     0     0     6     1
     6     9     6     4     9     8     1     4
     0     9     2     4     9     7     9     6
\end{lstlisting}
and calling
\begin{lstlisting}
0
[C,count]=MM(A,B,1,count)
\end{lstlisting}
results in the following operations.

The MM procedure computes first

\begin{equation}
row1=
\begin{bmatrix}
8  &   8  &   5  &    0 \\
     4  &   8 &    6  &   5  \\
     2   &  1  &   6 &    7  \\ 
     2  &   7  &   3  &  7  \\  
\end{bmatrix} +
    0.0001  
\begin{bmatrix}     0  &   4  &   8  &    0 \\
          0  &   3   &  7  &   5 \\
         3   &  3   &  7  &   1 \\
          0  &   3  &   1   &  4 \\
    \end{bmatrix}\label{eq1}
\end{equation}

\begin{equation}
col1=
\begin{bmatrix}
     4   &  1 &    4  &   4     \\
     2   &  5 &    8  &   1     \\
     4   &  2 &    8  &   9     \\
     6   &  3 &    6  &   9     \\
\end{bmatrix} +
    10000
\begin{bmatrix}  
     1  &   5 &    6 &    5   \\  
     0  &   0 &    5 &    4   \\  
     6  &   9 &    6 &    4   \\  
     0  &   9 &    2   &  4   \\
    \end{bmatrix}\label{eq2}
\end{equation}

\begin{lstlisting}
row1 = 
[   8.0, 8.0004, 5.0008,      0]
[   4.0, 8.0003, 6.0007, 5.0005]
[2.0003, 1.0003, 6.0007, 7.0001]
[   2.0, 7.0003, 3.0001, 7.0004]
\end{lstlisting}
and
\begin{lstlisting}
col1 =
[10004.0, 50001.0, 60004.0, 50004.0]
[    2.0,     5.0, 50008.0, 40001.0]
[60004.0, 90002.0, 60008.0, 40009.0]
[    6.0, 90003.0, 20006.0, 40009.0]
\end{lstlisting}

and it calls recursively 

\begin{lstlisting}
MM(row1,col1,2,count) 
\end{lstlisting}

to compute 

\begin{equation}
row1=
\begin{bmatrix}
 8.0 & 8.0004 \\
  4.0 & 8.0003 
\end{bmatrix} +
    0.00000001
\begin{bmatrix}  
5.0008 &  0 \\
6.0007 & 5.0005
    \end{bmatrix}\label{eq3}
\end{equation}

\begin{equation}
col1=
\begin{bmatrix}
10004.0 & 50001.0 \\
    2.0 &  5.0 \\
\end{bmatrix} +
    100000000
\begin{bmatrix}  
60004.0 &90002.0 \\
    6.0 & 90003.0 \\
    \end{bmatrix} \label{eq4}
\end{equation}

\begin{lstlisting}
row1 =
[8.000000050008,         8.0004]
[4.000000060007, 8.000300050005]
col1 =
[6000400010004.0, 9000200050001.0]
[    600000002.0, 9000300000005.0]
\end{lstlisting}
Then it calls again recursively 
\begin{lstlisting}
MM(row1,col1,3,count) 
\end{lstlisting}

to compute again
\begin{equation}
\begin{aligned}
row1 & =8.000000050008 \\ 
         &+ 0.0000000000000001*8.0004 \\
&=
8.00000005000800080004  \label{eq5}
\end{aligned}
\end{equation}
\begin{equation}
\begin{aligned}
col1 & =6000400010004.0  \\
      &+ 10000000000000000*600000002.0 \\
& = 6000000020006000400010004.0 \label{eq6}
\end{aligned}
\end{equation}
and since the dimension of blocks is one by one now, it returns
\begin{equation}
\begin{aligned}
mod(floor(r& ow1*col1),10000) = \\ 
mod(floor(&8.00000005000800080004 * \\
               &6000000020006000400010004.0),10000)=\\
mod(floor(& 48000000460096009000780116.009301),10000)=\\
mod(& 48000000460096009000780116,10000)=116 \label{eq7}
\end{aligned}
\end{equation}
which is returned as the first value of the result.
By traveling through the branches of the recursion tree, 
we finally obtain the result 
\begin{lstlisting}
C=
[116, 130, 204, 133, 221, 171, 109, 120]
[128, 179, 225, 183, 242, 218, 145, 176]
[121, 127, 183, 185, 151, 155,  67, 135]
[ 82, 109, 159, 137, 133, 140, 118, 138]
[ 97, 118, 205, 186, 159, 133,  90, 122]
[146, 219, 241, 230, 260, 239, 167, 202]
[113, 193, 244, 214, 227, 195, 187, 198]
[ 60, 156,  88,  88, 171, 136, 107,  94]
\end{lstlisting}
which is equal to the MATLAB result of $A*B$, as measured using Euclidean norm in MATLAB 
\begin{lstlisting}
norm(A*B-C,2)=0
\end{lstlisting}
We have obtained this result with  
\begin{lstlisting}
counter = 832
\end{lstlisting}
floating-point operations.
The processing of an 8x8 matrix takes 832 operations (438 additions and multiplications of numbers, plus 384 floor and modulo operations) versus $8*8*8*2=1024$ additions and multiplications of the Bin\'et algorithm.

For larger matrices, we need to set number of digits processed by \emph{vpa} library exactly, e.g.
\begin{lstlisting}
digits(128)
\end{lstlisting}
allows to multiply $32\times 32$ matrices, assuming 4 digits accuracy.

In general, for multiplication of two matrices of size $N \times N$ we have $n=log_2(N)$ recursive calls, and the arbitrary precision arithmetic emulator \emph{vpa} uses $O(N)$ digits to run our algorithm.. 
Thus, to process matrices of dimension $64 \times 64$ with 4 digits, we need the number of \emph{vpa} digits to be set to $4*64=256$.

We can employ this MATLAB code to count operations for larger matrices (see Table \ref{tab1}).
We can read from this table that the exact computational cost is $(4log_2(N)+1)N^2$.

\begin{table}[h!]
\centering

\begin{tabular}{ |c| c |c|c|   }
\hline
$N$ & operations count & Bin\'et $2N^3$ & $C=count/N^2$ \\ 
\hline
4 & 144 & 128 & 9 \\
8 &  832 & 1024 & 13  \\
16 & 4,352 & 8192 & 17 \\
32 &  21,504 &  65,536  & 
21 \\
64 &  102,400 &  524288  & 
25 \\
128 & 475,136 &  4,194,304 & 
29 \\
256 &  2,162,688 &  33,554,432 & 
33 \\\
512 &  9,699,328 &  268,435,456 & 
37 \\
1024 &  42,991,616 &  2,147,483,648 & 
41 \\
2048 &  188,743,680 &  17,179,869,184 & 
45\\
4096 &  822,083,584 &  137,438,953,472 & 
49 \\
N=8192 &  3,556,769,792 &  549,755,813,888 & 
53 \\
\hline
\end{tabular}
\caption{Computational cost (including additions, multiplications, as well as modulo and floor operations) as a function of matrix size. Comparison with Bin\'et algorithm cost. The coefficient $C=4*log_2(N)+1$ in front of $O(N^2)$ computational complexity.}
\label{tab1}
\end{table}

\section{\revision{Comparison of the numerical accuracy and execution times}}

\revision{In this section, we provide the estimation of the $\epsilon$ parameter to avoid round-off errors when computing modulo and floor operations.
We assume that $a_{ij}\leq A_{max}$ for $i,j=1,...,N$, and $b_{ij}\leq B_{max}$ for $i,j=1,...,N$.
We notice that
\begin{equation}
C_{ij} = \floor{\left(\sum_{j=1,...,N}A_{ik} \epsilon^k\right)\left(\sum_{k=1,...,N}B_{kj}\epsilon^{-k}\right)} \mod \frac{1}{\epsilon}
\end{equation}
The product
\begin{equation}
C_{ij} = \sum_{n=1,...,N;m=1,...,N}A_{in}B_{mj}\epsilon^{n-m} =
\sum_{k=1,...,N}A_{ik}B_{kj}+\sum_{n=1,...,N;m=1,...,N;n \neq m}A_{ik}B_{kj}\epsilon^{n-m}
\end{equation}
The first term can be bounded by $A_{max}B_{max}N$. We define the $\epsilon$ as
\begin{equation}
\epsilon < \frac{1}{10^{\floor{log_10(A_{max}B_{max}N)+1}}}
\end{equation}
We round it up to the powers of 10, and we add one extra factor of $10$ to leave the margin for rounding error and to handle potential additive noise.}

\revision{This observation leads to the optimized tensorial version of the fast matrix-matrix multiplication that can be pre-generated for a given matrix size.}

\begin{lstlisting}
digits(floor(N*log(maxnumber)/log(2));
eps=vpa(10)^(2*d+2+floor(log(N)));
eps=1/eps;
epsilons=[eps eps^2 eps^3 eps^4 eps^5 eps^6 eps^7 eps^8];
epsilons_inv = vpa(1.0./epsilons);
epsilons_mat = [epsilons; epsilons; epsilons; epsilons; epsilons; epsilons; epsilons; epsilons];
epsilons_inv_mat = [epsilons_inv; epsilons_inv; epsilons_inv; epsilons_inv; epsilons_inv; epsilons_inv; epsilons_inv; epsilons_inv ];
Ascaled = transpose(A.*epsilons_mat);
Asum(1:N) = sum(Ascaled(1:N,:));
Bscaled = B.*transpose(epsilons_inv\_mat);
Bsum(1:N) = sum(Bscaled(:,1:N));
C=transpose(Asum)*Bsum;
C=mod(floor(C),1/eps);
\end{lstlisting}

\revision{We set up the number of digits to be processed by \emph{vpa} as $N\times log_2(2N10^d)$, where $N$ is the matrix size, and $d$ is the number of decimal digits on the input matrices. They are collected in Table \ref{tab:digits}.
\begin{table}
\begin{tabular}{ |c| c|c|c|c|c|c|c|c|c|c|c|c|   }
\hline
$N$ & 2 & 4 & 8 & 16 & 32& 48 & 64 & 96 & 128 & 144 & 192  \\
\hline
$digits$ & 10 & 25 & 58 & 133 & 298 & 475 & 660 & 1047 & 1449 & 2286 & 3154  \\
\hline
\end{tabular}
\caption{Number of \emph{vpa} digits to be processed when multiplying matrices of different dimension $N$.}
\label{tab:digits}
\end{table}
Notice that the multiplication of matrix $A$ by a matrix with $\epsilon^k$ values is performed element-wise using the $.*$ MATLAB operator, which has a quadratic cost. Similarly, multiplications of matrix $B$ by a matrix with $\epsilon^{-k}$ values have a quadratic cost.
Next, the summation of rows and columns in matrices $A$ and $B$, respectively, has a quadratic cost as well.
We get two vectors $Asum$ and $Bsum$, and we obtain the resulting matrix C by performing the rank1-update operation having the quadratic cost as well, followed by the floor and modulo operations. Also, notice that we operate with long-precision arithmetics using the \emph{vpa} library.}

\revision{We generate these routines in MATLAB. 
To make a fair comparison, we have to take into account that our MATLAB routine is processing rows and columns of matrices in MATLAB, while the classical star * operation in MATLAB calls highly optimized C code to multiply these matrices.
Our routine is not low-level optimized, and thus to make the comparison fair (without spending a long time on optimizing the C code) we compare against the high-level MATLAB implementation of Binet algorithm, processing rows and columns}
\begin{lstlisting}
for i=1:N
  for j=1:N
    C(i,j)=A(i,1:N)*B(1:N,j);
  end
end
\end{lstlisting}
\revision{Our algorithm takes two vectors \emph{Asum} and \emph{Bsum} and computes rank1-update, which for ${\cal O}(N)$ input performs ${\cal O}(N^2)$ computations. Thus, the RAM-to-processor register communication is on order $N$ smaller than the performed computations, which are of order $N^2$. The Binet algorithm does not use \emph{vpa}, while our algorithm uses the long precision library.}

\begin{figure}
\includegraphics[width=0.7\textwidth]{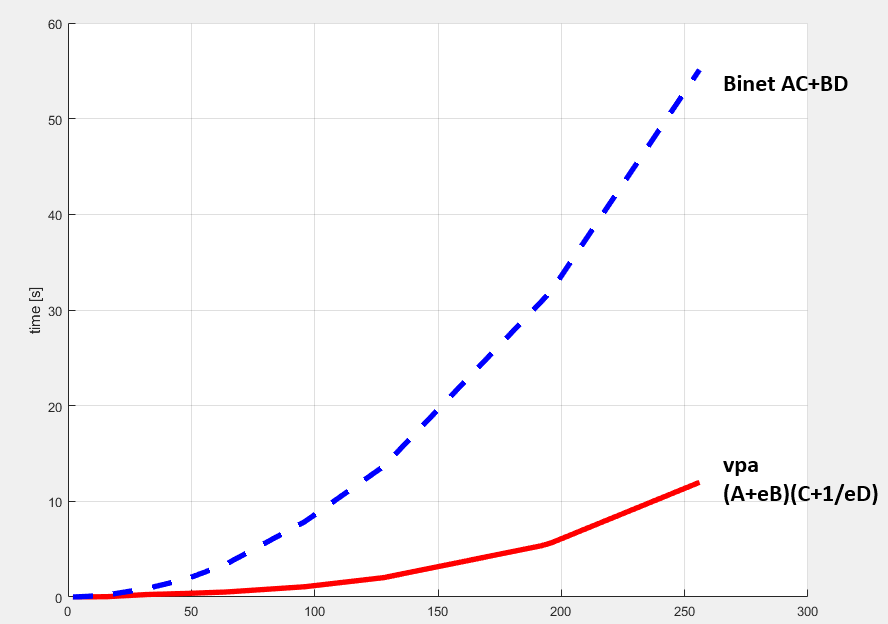}
\caption{Comparison of the MATLAB implementations of the matrix-by-matrix multiplication algorithm using long precision arithmetics with vpa and the classical recursive Bin\'et algorithm. Each algorithm has been executed 5 times.}
\label{fig:MATLAB}
\end{figure}

\revision{We have executed the experimental verification of the $A*B$ multiplication algorithm using the optimized code for growing dimensions of $A$ and $B$ for $N=2,4,8,16,32,64,96,128,144,192$ for the MATLAB implementation (see Figure \ref{fig:MATLAB}), processing integer matrices. In all the cases, MATLAB returned
$norm(C-A*B,2)$ equal to 0. We conclude that setting $\epsilon$ based on the bound of the maximum sum of products of rows and columns results in the numerical stability of the algorithm.}

\section{Conclusions}

We have proposed the algorithm that performs matrix-by-matrix multiplications in $O(N^2log_2N)$ computational complexity.
This computational complexity is bounded by $O(N^{2+\delta})$ for arbitrary small $\delta>0$.
The exact computational cost (the exact number of operations) of our algorithm is $(4log_2(N)+1)N^2$.
\revision{While deriving our algorithm, we assumed that we could add and multiply with $O(1)$ computational complexity numbers with a variable number of digits.
This assumption was based on the following rationale. Modern processor architectures like AVX-512 allow for ${\cal O}(1)$ processing of numbers up to 512 bits. Using this architecture, it is possible to process small matrices in ${\cal O}(1)$ time.
For larger matrices, it is possible to construct a multi-core processor that will process numbers using a parallel version
of FFT-based multiplication algorithms (like Schönhage-Strassen \cite{SS} or Fürer’s algorithm \cite{Furer}) or the Harvey–van der Hoeven algorithm with ${\cal O}(\log d)$ cost.}
We also assumed that we could perform modulo and floor operations on these numbers with $O(1)$ computational complexity.
We also assumed that the cost of accessing and storing matrix entries is zero.
We implemented our algorithm in MATLAB and Octave using \emph{vpa} variable precision arithmetic emulator. This emulator \emph{vpa} requires $O(N)$ digits to run our algorithm.
We have a simple algorithm with quadratic cost and a constant equal to 4 for the price of arbitrary precision arithmetic that can process small matrices efficiently.
One can hope that future computers will allow the storage of numbers with a variable number of digits and that these numbers could be processed with ${\cal O}(1)-{\cal O}(log d)$ computational complexity there.
In that case, our algorithm will be the winner.
Future work may involve the design of the parallel version of our algorithm and implementation on GPGPU using variable precision arithmetic.


\appendix

\section{\revision{Domain of algebraic complexity and the model of computations}}

\revision{The proposed algorithm involves additions $(a+b)$, multiplications $(a*b)$, inverse operation $(1/a)$, modulo operation $(a \; mod \; b)$, and floor $\floor{(a)}$ operations.
The modulo or floor operation is not defined over the commutative ring or over any field \cite{Book}.
To define a modulo operation, we must work outside of fields and rings and move to more general algebraic structures.
For this purpose, we can define the domain
\begin{equation}
N_Q  = \{(S,W)\in Z \times Z: 0\leq W < Q, Q \in Z \}
\end{equation}
($Q$ is introduced to prescribe the maximum number of digits of $W$, the fractional part).
Each element of $N_Q$ represents $val(S,W)=S+W/Q$. Here, $/$ represents a division of integers.
We introduce the following operations:
\begin{itemize}
\item division operation: $Z/Q = (\floor{Z/Q},Z \% Q)$ where $Z=(S_1Q+W_1)$ is an integer,
\item floor operation: $\floor{(S_1,W_1)} = S_1$,
\item modulo operation: $(S_1,W_1) \mod (S_2,W_2) = (M/Q,M \% Q)$ where $M = (S_1Q+W_1 ) \% (S_2Q+W_2)$
(here $(S_1Q+W_1 )$ and $(S_2Q+W_2 )$ are standard integers and $\%$ represents standard modulo operation over integers, and $/Q$ represents standard division of integers)
\item addition: $(S_1,W_1)+(S_2,W_2)=(S_1+S_2 +\floor{(W_1+W_2)/Q)}, (W_1+W_2) \% Q)$,
\item multiplication: $(S_1,W_1)*(S_2,W_2)=(\floor{Z/Q}, Z \% Q)$ where $Z=(S_1Q+W_1)(S_2Q+W_2)/Q$
(we introduce the scaled integers, like $S_1Q+W_1$  and $S_2Q+W_2$, and we multiply two integers to obtain Z. We scale them back to the original representation by dividing the integer by $Q$ and by computing the modulo of the integer, namely $Z/Q$ and a fractional part $Z\%Q$.
Here again, $\%$ stands for the standard modulo operation over integers)
\end{itemize}}

\revision{In our model of computation, we process pairs of integer numbers $(S,W)$, where $S$ and $W$ are represented as strings of digits.
One integer $S$ represents the digits „before the dot,” and the second integer $W$ represents the digits „after the dot.”
In our computational model, we assume that
\begin{itemize}
\item The cost of division, assuming $Q=2k$ is ${\cal O}(1)$ since it can be computed by bit shift and bitwise AND.
\item The module operation is
$(S_1,W_1) \mod (S_2,W_2) = (M/Q,M \& Q)$ where $M = S_1Q+W_1  \% S_2Q+W_2$.
The cost of modulo operation is ${\cal O}(1)$.
\item The floor operation is $\floor{(S_1,W_1)} = S_1$. The cost of floor operation is ${\cal O}(1)$.
\item  The addition operation is
$(S_1,W_1)+(S_2,W_2)=(S_1+S_2 +\floor{(W_1+W_2)/Q}, (W_1+W_2) \% Q) $
The cost of addiction is ${\cal O}(1)$
\item The multiplication is $(S_1,W_1)*(S_2,W_2)=(\floor{Z/Q}, Z \% Q)$ where $Z=(S_1Q+W_1)(S_2Q+W_2)/Q$.
The cost of multiplication is ${\cal O}(1)$ for small fixed-size integers that can be processed by modern processors (for small matrices) or for long integers that can be processed by modern AVX512 architecture. The cost of multiplication is ${\cal O}(M(d))$ for integers of size $d$ where $d$ is the number of digits and $M(d)$ is the cost of multiplication of long integers.
\end{itemize}
The complexity of long integer multiplication depends on the size of the manipulated numbers; see \cite{Harvey2}.
How can $M(d)$ be estimated? Single core processors enable multiplying two integers with $d$ digits using one of the fast multiplication algorithms, e.g., Karatsuba ${\cal O}(M(d))=O(d^{1.58})$ \cite{c12} or Harvey–van der Hoeven algorithm and $O(M(d))=O(d \log d)$ \cite{Hoeven}).
For multiple core processors, it is possible to multiply long integers with $d$ digits in time ${\cal O}(\log d)$ using the parallel version of FFT-based multiplication algorithms (like Schönhage–Strassen \cite{SS} or Fürer’s algorithm \cite{Furer}) and possibly the parallel version of Harvey–van der Hoeven \cite{Hoeven}.
Thus, the cost of multiplication of two matrices of size $N \times N$ using a multicore processor that can process $N$ digits in time ${\cal O}(\log N)$ is ${\cal O}(N^2(\log N)(\log N))$.}


\begin{thebibliography}{8}

\bibitem{c1}
Josh Alman and Virginia Vassilevska Williams. 2021. A refined laser method and
faster matrix multiplication. In Proceedings of the Thirty-Second Annual ACMSIAM
Symposium on Discrete Algorithms (Virtual Event, Virginia) (SODA ’21).
Society for Industrial and Applied Mathematics, USA, 522–539.
\bibitem{c2} E. Anderson, Z. Bai, C. Bischof, L. S. Blackford, J. Demmel, J. Dongarra,
J. Du Croz, A. Greenbaum, S. Hammarling, A. McKenney, and D.
Sorensen. 1999. LAPACK Users’ Guide (third ed.). Society for Industrial
and Applied Mathematics. https://doi.org/10.1137/1.9780898719604
arXiv:https://epubs.siam.org/doi/pdf/10.1137/1.9780898719604
\bibitem{c3} Dario Bini, Milvio Capovani, Francesco Romani, and Grazia Lotti. 1979. $O(n^{2.7799})$
complexity for $n\times n$ approximate matrix multiplication. Inform. Process. Lett. 8,
5 (1979), 234–235. https://doi.org/10.1016/0020-0190(79)90113-3
\bibitem{c4} H. Cohn, R. Kleinberg, B. Szegedy, and C. Umans. 2005. Group-theoretic algorithms
for matrix multiplication. In 46th Annual IEEE Symposium on Foundations
of Computer Science (FOCS’05). 379–388. https://doi.org/10.1109/SFCS.2005.39
\bibitem{c5} Henry Cohn and Christopher Umans. 2013. Fast matrix multiplication using
coherent configurations. Proceedings of the Twenty-Fourth Annual ACM-SIAM Symposium on Discrete Algorithms. 1074–1087. https://doi.org/10.1137/1.9781611973105.77
arXiv:https://epubs.siam.org/doi/pdf/10.1137/1.9781611973105.77
\bibitem{c6} H. Cohn and C. Umans. 2003. A group-theoretic approach to fast matrix multiplication.
In 44th Annual IEEE Symposium on Foundations of Computer Science,
2003. Proceedings. 438–449. https://doi.org/10.1109/SFCS.2003.1238217
\bibitem{c7} D. Coppersmith and S. Winograd. 1982. On the Asymptotic Complexity of Matrix
Multiplication. SIAM J. Comput. 11, 3 (1982), 472–492. https://doi.org/10.1137/
0211038 arXiv:https://doi.org/10.1137/0211038
\bibitem{c8} D. Coppersmith and S. Winograd. 1987. Matrix multiplication via arithmetic
progressions. In Proceedings of the Nineteenth Annual ACM Symposium on Theory
of Computing (New York, New York, USA) (STOC ’87). Association for Computing
Machinery, New York, NY, USA, 1–6. https://doi.org/10.1145/28395.28396
\bibitem{c9} Don Coppersmith and Shmuel Winograd. 1990. Matrix multiplication via arithmetic
progressions. Journal of Symbolic Computation 9, 3 (1990), 251–280.
https://doi.org/10.1016/S0747-7171(08)80013-2 Computational algebraic complexity
editorial.
\bibitem{c10} A. M. Davie and A. J. Stothers. 2013. Improved bound for complexity of matrix
multiplication. Proceedings of the Royal Society of Edinburgh: Section A Mathematics
143, 2 (2013), 351–369. https://doi.org/10.1017/S0308210511001648
\bibitem{Nature}
Alhussein Fawzi, Matej Balog, Aja Huang, Thomas Hubert, Bernardino Romera-
Paredes, Mohammadamin Barekatain, Alexander Novikov, Francisco J. R. Ruiz,
Julian Schrittwieser, Grzegorz Swirszcz, David Silver, Demis Hassabis, and Pushmeet
Kohli. 2022. Discovering faster matrix multiplication algorithms with
reinforcement learning. Nature 610 (2022), 47–53.
\bibitem{c11} Jianyu Huang, Chenhan D. Yu, and Robert A. van de Geijn. 2020. Strassen’s
Algorithm Reloaded on GPUs. ACM Trans. Math. Softw. 46, 1, Article 1 (March
2020), 22 pages. https://doi.org/10.1145/3372419
\bibitem{c12} A. Karatsuba and Yu. Ofman. 1962. Multiplication of Many-Digital Numbers by
Automatic Computers. Proceedings of the USSR Academy of Sciences 145 (1962),
293–294.
\bibitem{c13} François Le Gall. 2014. Powers of tensors and fast matrix multiplication. In
Proceedings of the 39th International Symposium on Symbolic and Algebraic Computation
(Kobe, Japan) (ISSAC ’14). Association for Computing Machinery, New
York, NY, USA, 296–303. https://doi.org/10.1145/2608628.2608664
\bibitem{c14} John O’Connor and Edmunc Robertson. 2024. Jacques Philippe Marie Bin\'et,
MacTutor History of Mathematics Archive. University of St Andrews. Mathshistory.
st-andrews.ac.uk
\bibitem{c15} V. Ya. Pan. 1978. Strassen’s algorithm is not optimal trilinear technique of
aggregating, uniting and canceling for constructing fast algorithms for matrix
operations. In 19th Annual Symposium on Foundations of Computer Science (sfcs
1978). 166–176. https://doi.org/10.1109/SFCS.1978.34
\bibitem{c16} V. Ya. Pan. 1980. New Fast Algorithms for Matrix Operations. SIAM
J. Comput. 9, 2 (1980), 321–342. https://doi.org/10.1137/0209027
arXiv:https://doi.org/10.1137/0209027
\bibitem{c17} Francesco Romani. 1982. Some Properties of Disjoint Sums of Tensors Related to
Matrix Multiplication. SIAM J. Comput. 11, 2 (1982), 263–267. https://doi.org/10.
1137/0211020 arXiv:https://doi.org/10.1137/0211020
\bibitem{c18} A. Schönhage. 1981. Partial and Total Matrix Multiplication. SIAM
J. Comput. 10, 3 (1981), 434–455. https://doi.org/10.1137/0210032
arXiv:https://doi.org/10.1137/0210032
\bibitem{SS} Arnold Schönhage, Volker Strassen. 1971. Multiplikation großer Zahlen.
Computing 7 (1971), 281–292. https://doi.org/10.1007/BF02242355
\bibitem{c19} Fengguang Song, Jack J. Dongarra, and Shirley V. Moore. 2006. Experiments with
Strassen’s Algorithm: From Sequential to Parallel. https://api.semanticscholar.
org/CorpusID:15137373
\bibitem{c20} Volker Strassen. 1969. Numerishe Mathematics 13 (1969), 354–356. https://doi.
org/10.1007/BF02165411
\bibitem{c21} V. Strassen. 1986. The asymptotic spectrum of tensors and the exponent of matrix
multiplication. In 27th Annual Symposium on Foundations of Computer Science
(sfcs 1986). 49–54. https://doi.org/10.1109/SFCS.1986.52
\bibitem{c22} Wikipedia. 2024. Computational complexity of matrix multiplication. Retrieved
October 21, 2024 from https://en.wikipedia.org/wiki/Computational\_complexity\_
of\_matrix\_multiplication
\bibitem{c23} Virginia Vassilevska Williams. 2012. Multiplying matrices faster than
Coppersmith-Winograd. In Proceedings of the Forty-Fourth Annual ACM Symposium
on Theory of Computing (New York, New York, USA) (STOC ’12). Association
for Computing Machinery, New York, NY, USA, 887–898. https:
//doi.org/10.1145/2213977.2214056
\bibitem{vpa}
https://www.mathworks.com/help/symbolic/sym.vpa.html
\bibitem{Hoeven}
\revision{David Harvey, Joris van der Hoeven. Integer multiplication in time ${\cal O}(n \log n)$. Annals of Mathematics (2021)
10.4007/annals.2021.193.2.4}
\bibitem{Furer}
\revision{Martin Fürer, Faster Integer Multiplication, SIAM Journal on Computing 39(3) (2009) 979–1005.}
\bibitem{Williams1}
\revision{Virginia Vassilevska Williams, Multiplying matrices in ${\cal O}(n^{2.373})$ time 
https://people.csail.mit.edu/virgi/matrixmult-f.pdf}
\bibitem{Williams2}
\revision{Josh Alma, Virginia Vassilevska William, Limits on All Known (and Some Unknown) Approaches to Matrix Multiplication (2018)
https://people.csail.mit.edu/virgi/matmultlimits.pdf}
\bibitem{Book}
\revision{Peter Bürgisser , Michael Clausen , Mohammad Amin Shokrollahi, Algebraic Complexity Theory, Springer Nature, 1997.}
\bibitem{Harvey2}
\revision{David Harvey, Joris van der Hoeven, On the complexity of integer matrix multiplication, Journal of Symbolic Computation, 89 (2018), 1-8.}
\end{thebibliography}

\end{document}